\documentstyle[epsf]{mn}

\title[Neutral hydrogen at high redshifts]
      {Neutral hydrogen at high redshifts as a probe of structure
	formation -- III.  Radio maps from N-Body simulations.}
\author[J.S.Bagla, Biman Nath and T.Padmanabhan]
       {J.S.Bagla\thanks{Present Address: Institute of Astronomy, Madingley Road, Cambridge CB3 0HA, U.K. \ \   e-mail : jasjeet@ast.cam.ac.uk},
	Biman Nath\thanks{E-mail: biman@iucaa.ernet.in} and   
        T.Padmanabhan\thanks{paddy@iucaa.ernet.in}\\
	Inter-University Centre for Astronomy and Astrophysics, Post Bag 4,
        Ganeshkhind, Pune 411 007, INDIA}

\date{IUCAA Preprint 40/96}

\pagerange{\pageref{firstpage}--\pageref{lastpage}}
\pubyear{1997}

\begin{document}
\label{firstpage}

\maketitle

\begin{abstract}
Large inhomogeneities in neutral hydrogen in the universe can be
detected at redshifts $z \leq 10$ using the redshifted $21$cm line
emission.  We use cosmological N-Body simulations for dark matter and a
simple model for baryonic collapse to estimate the signal expected
from structures like proto-clusters of galaxies at high redshifts.  We
study~: (i) the standard CDM model, (ii) a modified CDM model with less power
at small scales, and (iii) a $\Lambda+$CDM model in a universe with
$\Omega_0 + \Omega_\Lambda = 1$.  We show that it should be
possible for the next generation radio telescopes to detect such
structures at the redshift $3.34$ with an integration of about $100$
hours.  We also discuss possible schemes for enhancing signal to noise
ratio to detect proto-condensates at high redshifts.
\end{abstract}

\begin{keywords}
Galaxies : formation -- cosmology : theory -- early Universe,
large scale structure of the Universe
\end{keywords}

\section{Introduction}

It is generally believed that large scale structures like galaxies and
clusters of galaxies formed from small initial inhomogeneities via
gravitational collapse.  One implication of this picture is a distinct
epoch when structures like proto-galaxies and proto-clusters decoupled 
from the largely homogeneous universe.  Present observations suggest that 
this epoch is around $z\sim 5$ for galaxies.  Observations of
structures in this stage of formation, if made, can be 
a very powerful constraint on the models of structure formation.  Such
observations will also improve our understanding of the process of
structure formation. 

Sunyaev and Zel'dovich (1972) [also see Sunyaev and Zel'dovich (1975)]
pointed out that the formation of first structures may be probed by
observing the redshifted $21$cm line emitted by the neutral hydrogen in
these structures.  Several searches have been made to look for such
structures at high redshifts.  In absence of a detection, these
searches have only been able to put limits on the mass of neutral
hydrogen present in clumped form.  For a summary of these surveys see
Wieringa, de Bruyn and Katgert (1992) and references cited in that paper. 

The Giant Meter-wave Radio Telescope (GMRT) presently being constructed
in India should be able to improve the observational situation
considerably as regards the detection and study of proto condensates
containing neutral hydrogen \cite{swarup84}.  The GMRT will be able to
probe the redshifted $21$cm line from three epochs centred at
$z=3.34$, $5.1$ and $8.5$.  In this paper we will discuss the
possibility of detection at the two lower redshifts.

Subramanian and Padmanabhan (1993) have computed the expected flux at
these redshifts for some models of structure formation.  They used the
Press-Schechter formalism \cite{ps75} to compute the expected number
densities of proto-clusters in the CDM and HDM models.  In a later paper
\cite{nhline} they computed line profiles assuming the proto-clusters
to be spherically symmetric.  These studies suggest that it should be
possible to detect proto-clusters in the standard CDM model using the
GMRT with $10$ to $20$ hours of observations.

In this paper we will use N-Body simulations to follow gravitational
collapse of dark matter and some simple approximations for
estimating the collapsed and neutral fraction in dense regions to
construct ``radio maps'' with same specifications as the GMRT.  We
will study these maps and suggest simple methods to optimise the signal
to noise ratio.   

Some authors have studied the distribution of neutral hydrogen at high
redshifts using simulations that include gas dynamics, ionisation and
other astrophysical processes.  Most of these studies focus on small scale
variations in the distribution of neutral hydrogen.  [e.g. see
\cite{davwein}]  However, the synthesised beam for the central square
of the GMRT includes a large comoving volume in each pixel (approximately
$8(h^{-1}$Mpc$)^3$) and so the details of physical processes operating
at small scales can be ignored to a large extent.  We can also ignore
the differences in distribution of baryons and dark matter at small
scales.  Further, as we are interested in the neutral fraction at two
epochs, we can choose to ignore the physical processes responsible for
its evolution.  This simplifies the problem to a large extent and we
should be able to get meaningful estimates of the signal strength
without a detailed treatment of baryons and astrophysical processes.

A brief discussion of the expected
physical conditions at the epochs of interest is given below.  These
are the guiding considerations in choosing the simplifying assumptions
for constructing the radio maps.
\begin{itemize}
\item $z=3.34$ :  Observations show that the inter-galactic medium
(IGM) is completely ionised \cite{gptest} at this redshift.  Numerical
simulations suggest that nearly all of the neutral hydrogen is in high
density, radiatively cooled objects {\cite{davwein}}.  Observations
show that a large fraction of mass in the damped lyman-$\alpha$
absorption systems (DLAS), believed to be progenitors of present day
galaxies, is in form of neutral hydrogen {\cite{dlas_omnh1}}.
Observations also show that the spin temperature of gas in DLAS, the
relevant quantity for the $21$cm transition, is much higher than the
temperature of the background radiation at this epoch.  It can be
shown that, in such conditions, emission is the dominant mechanism and
{\it the total energy emitted by radiators in such a state does not
depend on the spin temperature}\/ \cite{screes}. 

\item  $z=5.1$ : We have very little information about the universe at
this redshift.  It is known that the intergalactic medium at redshifts
$z< 5$ is fully ionised \cite{gptest}.  It is believed that the
process of ionisation is initiated by the first luminous structures
in the universe.  First structures like galaxies and quasars can form
after $z\simeq 5-6$ in most models that satisfy other observational
bounds.  There are three possible scenarios that deserve mention.
\begin{itemize}
\item  One possibility is that the universe was already reionised
by $z=5.1$.  In this case the neutral hydrogen would be confined to dense
regions like proto-galaxies and the intervening regions will be
completely, or mostly, ionised.  

\item  If the universe has been reheated by first luminous objects,
but not reionised, then the spin temperature will be much higher that
the temperature of the background radiation.  A patchy reheating could
lead to fluctuations in the $21$cm emission at large scales \cite{tomo}.  

\item  If the universe is neither reheated not reionised by $z=5.1$
and the first luminous objects form around this epoch, then the spin
temperature will be comparable to the temperature of the background
radiation for models with $\Omega_b \leq 0.1$ \cite{screes}.
\end{itemize}

We will focus on the second scenario (for $z=5.1$) and carry out all our 
calculations assuming that the universe is largely neutral and the spin
temperature is much larger than the temperature of the radiation
background.  Fluctuations in spin temperature at large scales
\cite{tomo} introduce uncertainty in the results as the region of
interest may not have been reheated and hence the $21$cm radiation
will be much less than anticipated.

\end{itemize}

\section{Generating Radio Maps}

We will now outline the method that is used to construct radio maps
by combining the distribution of dark matter obtained from N-Body
simulations with approximations for the neutral fraction of gas.
Radio observations with an interferometer give the flux of radiation
coming from a given direction in the sky in a range of frequencies.
This range of frequencies is subdivided into small equal intervals
(channels).  Within the field of view (primary beam) we can
differentiate between flux from directions separated by an angle equal
to the resolution of the interferometer (synthesised beam).  Therefore
we can arrange the information obtained from such an observation in
terms of a radio map for each frequency channel.  The angular
resolution, or the pixel size, with which the map is constructed is
given by the synthesised beam of the interferometer.  For the purpose
of generating ``radio maps'' from simulations, we will choose
the parameters like pixel size and channel width to be same as that
used by the GMRT central array \cite{swarup84}.  (The central array
consists of $12$ antennas of diameter $45$m spread over a region
of $1$km $\times$ $1$km.  The telescopes are scattered randomly within
this region.)

Earlier estimates of expected flux from high redshift objects have
shown proto-clusters to be the most promising source \cite{nhI} of
$21$cm radiation.  Observations of high redshift objects like quasars
suggest that the intergalactic medium is completely ionised at
redshifts $z < 4$.  This implies that the only source of neutral
hydrogen will be dense clouds inside galaxies.  These clouds are
sufficiently dense so that absorption in a thin layer near the surface
shields the inner regions and a large fraction of the gas in these
remains unionised.  Therefore, while computing the flux, we must take
into account the fact that the neutral hydrogen we observe resides in
dense clumps and shares the internal velocity dispersion of these.  We
take this fact into account by convolving the line profile with a
Gaussian of width $200$km$/$s, at the epoch of emission.  This
consideration is relevant only for the window at $z=3.34$ as in most
models galaxies have not formed by $z=5.1$.  Any proto-galaxies that
may exist are not expected to have any systematic velocity dispersion. 

A remaining uncertainty in computing the amount of neutral hydrogen
at high redshifts is the fraction of mass in galaxies at the epoch of
interest.  We will {\it reduce} the uncertainty due to this factor by 
considering only the regions with $\delta > 1$ in the numerical
simulations.  (Changing this threshold to $\delta > 3$ reduces the
peak signal by less than $15\%$.  Therefore the final result is not
very sensitive to this threshold.)  However, since this is a complex
issue, we will only 
parameterise our ignorance with a factor $f_{gal}$ which is the
fraction of mass in galaxies that can hold neutral hydrogen.  We will
use $f_{gal}=1$ but the results can be rescaled with any value.  In a
more detailed calculation, which can be done with N-Body simulations
that use a much larger number of particles, individual halos of mass
$M \ge 10^{10} M_\odot$ can be directly identified and the
uncertainty parametrised by $f_{gal}$  becomes irrelevant.  This can
not be done at present without reducing the physical size of the
simulation box by a significant amount - in which case we will not be
able to map all scales probed by the GMRT.

The radio map is generated from the N-Body data in the following manner:
\begin{itemize}
\item  Use the co-ordinates of the particle to fix the pixel to which the
contribution is to be added.  

\item  Use the radial component peculiar velocity of the particle to
compute the total redshift.  Convolve the line profile with a Gaussian
of width $200$km$/$s and add the contribution to the relevant
frequency channels.

\item For generating maps at redshifts $z<4$  impose a cutoff on
density contrast $\delta \geq 1$ to ensure that we are looking at
regions that can host galaxies.

\item  Repeat this process for all particles.
\end{itemize}
In the following subsections we will describe each of these components
in greater detail.  We begin by outlining the models of structure
formation that are used for this study.

\subsection{N-Body Simulations}

We used N-Body simulations of three models to generate the ``radio maps.''  
These were chosen from the family of CDM models and we used the
parametrised spectrum given by Efstathiou, Bond and White (1992).  
These models were normalised using the root mean square
amplitude of fluctuations in the temperature of the CMBR (Cosmic
Microwave Background Radiation) observed by COBE \cite{cobe4}. 

All the simulations were carried out using $128^3$ particles in a $128^3$
mesh.  The physical size of the box in each case was $128 h^{-1}$Mpc.
Thus the mass of each particle in these simulations equals $2.7 \times
10^{11} \Omega_0 M_\odot$. 

The following parameters were chosen for these models~:

\begin{figure}
\epsfxsize=3.25 true in \epsfbox[34 23 547 658]{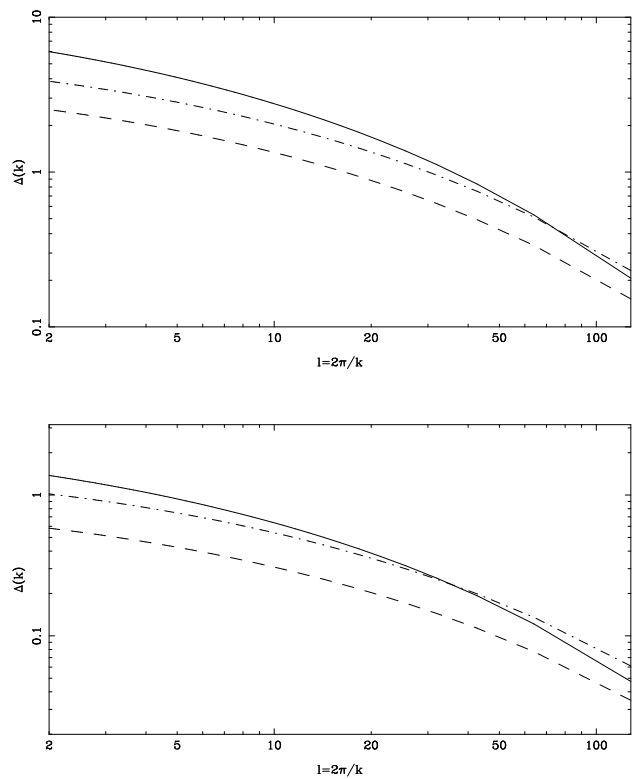}
\caption{This figure shows the power spectra for the three models being
discussed here.  The standard CDM model is shown using a thick line,
model~II (MDM) as a dashed line and model~III (LCDM) as a
dot-dashed line in these 
panels.  We have shown the linearly evolved power spectrum at the
present epoch [top panel] and also at the redshift $z=3.34$ [lower panel].
The different rate of growth for model~III (LCDM) in the
linear regime leads to 
an increase in the ratio of power in this model to the power in
model~I (sCDM) or model~II (MDM).}
\end{figure}
\begin{itemize}
\item  {\sl Model I } : The standard CDM model (sCDM). ($\Gamma = \Omega_{nr}
h = 0.5$, $\Omega_{nr} = \Omega_0 = 1$ and $h=0.5$.)  Normalisation with
fluctuations in the CMBR gives $\sigma(8 h^{-1}\hbox{Mpc},z=0) \simeq 1.2$
which is about twice the value implied by the abundance of rich
clusters. This model is a reference model in studies of structure
formation. 

\item {\sl Model II } :  A flatter version of the CDM model with
$\Gamma=0.3$, $\Omega_{nr}= 
\Omega_0 = 1$ and $h=0.5$.  The choice of $\Gamma$ for this model is
independent of the cosmological parameters.  This model has $\sigma(8
h^{-1}\hbox{Mpc},z=0) \simeq 0.6$ which is consistent with the observed
amplitude of fluctuations at this scale.  This model also predicts the
correct slope of the correlation function at large scales.  At large
scales this model is a close approximation for some ``flat'' versions
of CDM that can arise if a small but non-negligible fraction of mass
in the universe is contributed by relativistic dark matter. (Mixed
Dark Matter or MDM hereafter.)

\item  {\sl Model III } : A $\Lambda +$CDM model (LCDM).  ($\Gamma =
\Omega_{nr} h = 0.3$, 
$\Omega_{nr} = 0.6$, $\Omega_\Lambda = 0.4$ and $h=0.5$.)  This model has 
$\sigma(8 h^{-1}\hbox{Mpc},z=0) \simeq 1$ which is also consistent with
the observed abundance of rich clusters.  [The observed amplitude of
fluctuations at the scale of $8 h^{-1}\hbox{Mpc}$ scales approximately
as $\Omega_{nr}^{-0.6}$ \cite{sigma8}.]  This model satisfies most
observational constraints that are available for cosmological models
and models of structure formation. \cite{comments} (These constraints
rule out large regions of the parameter space and only 
a small region survives.  Therefore model~III is a good example from
the class of allowed models.)  This model was chosen for two reasons :
\begin{itemize}
\item  Growth of perturbations in a $\Lambda$ model slows down at late
epochs.  In other words, such a model has more power at early times in
comparison with an $\Omega_{nr}=1$ universe which has same level of
clustering at $z=0$.  Thus we have reason to expect higher signal for
models in a universe with nonzero $\Lambda$.

\item  The comoving volume enclosed in a given solid angle at high
redshifts is higher for a universe with nonzero $\Lambda$.  This implies
more emitters and hence a higher signal.  In a model with nonzero
cosmological constant the luminosity distance is also larger in
comparison with the corresponding distance in the Einstein-de Sitter
model for any given redshift.  However, the increase in comoving
volume mentioned above compensates for the increase in luminosity
distance in such models.  
\end{itemize}
We have shown the power spectra for these models in figure~1.  To
demonstrate the effect of different rate of growth in linear regime we
have also shown the same spectra at redshift $3.34$.
\end{itemize}

We would like to emphasise the following points regarding the choice of
models: It is possible to choose many variants of the above models and
repeat the analysis given in this paper.  One can choose to normalise
the power spectra in a different manner, e.g. by fixing the amplitude
of perturbations at $8 h^{-1}$Mpc rather than at very large scales
using COBE observations.  Other parameters like the Hubble's constant,
the cosmological constant, the density parameter and the primordial
spectral index can also be varied from the values used here.  We have
consciously restricted ourselves to a small set of example models for
the purpose of illustration.  A more detailed exercise will be
warranted if the GMRT succeeds in detecting neutral hydrogen at these
redshifts and a more meaningful comparison with models can be carried
out using real data.

\subsection{Evolution of Neutral Fraction}

Generating artificial radio maps requires, in principle, a detailed
knowledge of the distribution of baryons and the neutral fraction
etc.  However, as mentioned earlier in this paper, the comoving volume
enclosed in each pixel is very large and therefore we can get
reasonable estimates by assuming the distribution of baryons and dark
matter to be the same.  In the following discussion we will
demonstrate, using a simple model, that the assumption of a constant
neutral fraction at these scales is not very far from the truth.  We
will also show that our choice of $f_N=0.5$ for $z=3.34$ is not 
unrepresentative.  
We
will show that the neutral fraction of gas in galaxies is largely
independent of the depth of the potential well.  In the following
discussion we estimate the neutral fraction by using the model of
star formation in galaxies of Kauffmann, White and Guiderdoni (1993; 
hereafter KWG).  Here we will briefly summarise the relevant
features of their model and use these to estimate the evolution of
neutral fraction. 

In this model, dark matter halos are assumed to be
truncated singular isothermal spheres and it is assumed that the
temperature $T$ of the gas is given in terms of the circular velocity,
$V_c$, as, $T=35.9 \> (V_c/$km$/$s$)^2$ K.  The virial radius $r_v$ is
defined to be the radius within which the mean overdensity is $200$
({i.e.}, $r_v=0.1 H_0^{-1} (1+z)^{-3/2} V_c$).  The radius where the
cooling time of the gas is equal to the age of the universe is defined
as the cooling radius, $r_{cool}$.

Suppose that the fraction of the critical density that is in baryons
is $\Omega_b$ and $f_g$ is the fraction of the baryons in the form of
gas. The amount of cold gas inside the halo at time $t$ is given by the
amount of gas with cooling time $t_{cool} < t$. When $r_{cool} \gg r_v$, 
cooling is very rapid and the rate of increase of cold gas in the halo
is governed by the accretion rate of the halo, \cite{wf91}
\begin{equation}
\dot M_{inf} (V_c, z)=0.15 \> f_g \> \Omega_b \> V_c^3 \> G^{-1} \>.
\end{equation}
In the other limit, $r_{cool} \ll r_v$, the rate of inflow of cold gas
can be written as,
\begin{equation}
\dot M_{cool} (V_c, z)= 4 \pi \rho_g (r_{cool}) r_{cool}^2 
{dr_{cool} \over dt} \>,
\end{equation}
where $\rho_g (r)$ is the gas density at radius $r$. 

In the model of KWG, the rate at which cold gas settles inside the
halo is given by $\min (\dot M_{inf}, \dot M_{cool})$.  For the
cooling of the gas, we use the cooling function of Fall and Rees
(1985) for a primordial gas, since the metallicity is any way small
for gas at high redshift.

However, with the onset of star formation, the supernovae will begin
to heat the gas. Following KWG, we assume that the number of supernovae
per solar mass of stars formed is $\eta _{SN}=4 \times 10^{-3}$ M$_{\odot}
^{-1}$.  If each of the supernovae have a kinetic energy of the ejecta of
order $10^{51}$ erg, and a fraction $\epsilon$ of this energy is used
to heat the cold gas, then the rate of loss of cold gas to the hot
phase of the interstellar medium is, \cite{kwg}
\begin{equation}
\dot M_{reheat} = \epsilon {4 \over 5} { \dot M_{\ast} \eta _{SN} E_{SN}
\over V_c^2} \>.
\end{equation}
Here, $\dot M_{\ast}$ is the star formation rate, which is given in this
model as,
\begin{equation}
\dot M_{\ast} = \alpha {M_{cold} \over t_{dyn}} \>.
\end{equation}
Here $t_{dyn}$ is the dynamical time. The authors defined the dynamical
time as (where $\lambda \sim 0.05$ is the initial dimensionless spin
parameter of the gas),
\begin{equation}
t_{dyn}={r_{gal} \over V_{gal}}={(2 \lambda r_{vir} )^{3/2} \over
(G M_{gal})^{1/2} } \>.
\end{equation}

\begin{figure}
\epsfxsize=3.5 true in \epsfbox[18 144 592 718]{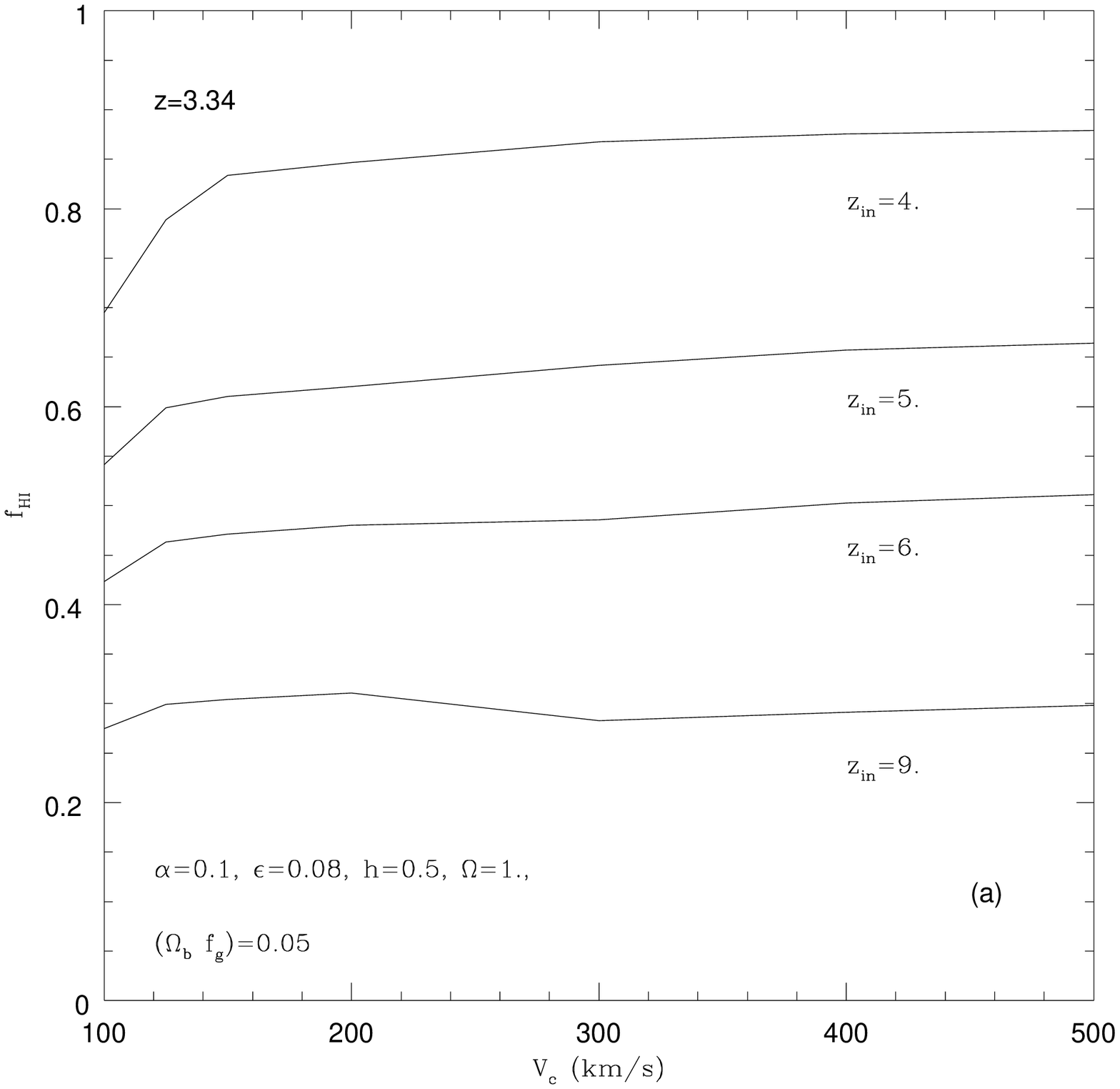}
\epsfxsize=3.5 true in \epsfbox[18 144 592 718]{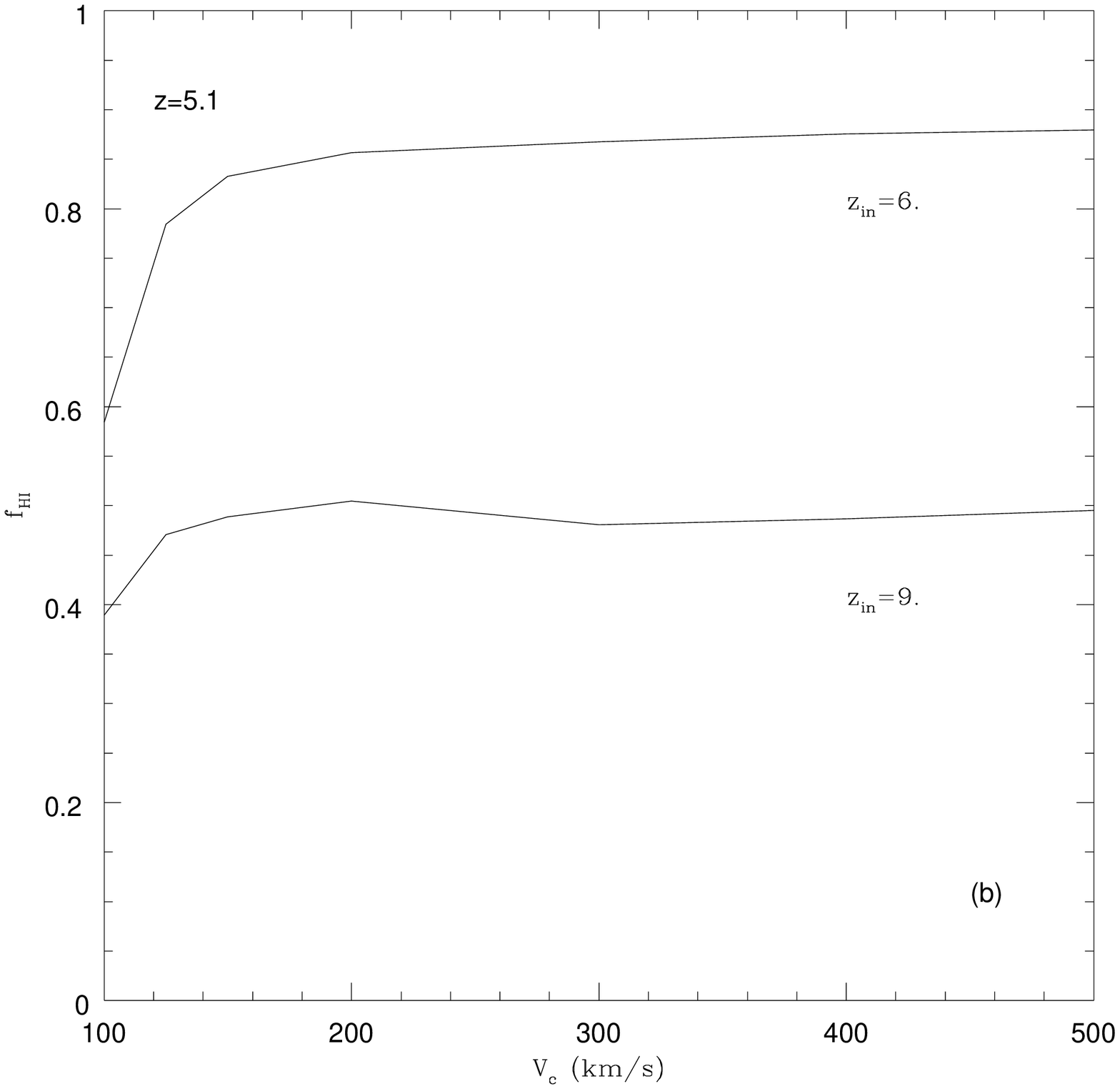}
\caption{The upper panel shows neutral fraction $f_{H~I}$ at $z=3.34$ 
against the circular velocity $V_c$ of the halo for $z_{in}=4, 5, 6, 9$.
We have assumed $\alpha=1$, $\epsilon=0.08$, $f_g \Omega_b=0.05$, $h=0.5$
and $\Omega=1$. Here, $f_g$ is the fraction of baryonic mass in the form of
gas. The curves do not strongly depend on the value of $f_g$.
The lower panel similarly plots the neutral fraction at
$z=5.1$ for $z_{in}=6,9$.}
\end{figure}

These equations govern the evolution of the amount of cold gas,
or, in other words, the neutral fraction $f_{HI}$, of gas in a halo of 
circular velocity $V_c$ at a given redshift $z$, for an assumed value of the
onset of inflow of gas $z_{in}$, and given the values of $\alpha$
and $\epsilon$. Gas is assumed to be fully ionised at $z_{in}$.
KWG estimated the values of $\alpha$
and $\epsilon$ from evolving the stellar population according to the
above model and comparing the mean luminosity and the cold gas content
of halos of $V_c=220$ km$/$s with the observed values for the Milky
Way [their Table 1]. For example, for $\Omega_0=1$, $\Omega_b=0.1$,
$\alpha=0.1$, and $\epsilon=0.08$. The values of $\alpha$
and $\epsilon$ are smaller for smaller values of $\Omega_b$. However,
we have found that the neutral fraction is not very
sensitive to the values of $\alpha$ and $\epsilon$, but rather to 
$z_{in}$. 

We plot in figure 2 the neutral fraction $f_{HI}$ at $z=3.34$ for
$z_{in}=4, 5, 6, 9$, as functions of the rotational velocity $V_c$. It
is seen that the neutral fraction is not sensitive to the value of
$V_c$, for $V_c > 150$ km$/$ s.  We will, therefore, use a
constant value of $f_n$ for all potential wells.  We also show the
values of $f_{HI}$ at $z=5.1$ for $z_{in}=6, 9$ as functions of $V_c$
in figure~2.  The neutral fraction $f_n$ does not depend very strongly
on the value of $f_g \Omega_b$.  In the following discussion we will
use $f_n=0.5$.

\subsection{Redshift Space Projection}

Simulations output the data in form of positions and velocities of
particles whereas radio observations detect the flux as a function of
angular position and frequency.  In this subsection we will outline the
method used for computing the observed central frequency around which the
$21$cm radiation from hydrogen atoms represented by a given N-Body
particle is observed.  

The peculiar velocity of an N-Body particle is given by
\begin{equation}
{\bf v}_p = a \dot{a} \frac{d{\bf x}}{da} = H_0 a^2\frac{H(a)}{H_0}
\frac{d{\bf x}}{da}  
\end{equation}
We only need the radial component of velocity for computing the redshift.
As we are dealing with particles at very high redshifts we can align one
axis of the simulation box along the line of sight and use that component
of velocity for computing the recession velocity.  The ratio of $H/H_0$ is
given by 
\begin{equation}
\frac{H(a)}{H_0} = \frac{\Omega_{nr}}{a^3} + \Omega_\Lambda
\end{equation}
for the models of interest.  The effective redshift results from a
combination of the Hubble recession and the peculiar velocity.
\begin{equation}
1 + z_{tot} = (1 + z_{hub})(1 + z_{pec})
\end{equation}
The average $z_{hub}$ for the simulation box is $1/a -1$.  Redshift due to
peculiar velocity is given by $z_{pec} = v_p/c$ where $c$ is the speed of
light. 

\subsection{Flux}

If the spin temperature of the atoms emitting $21$cm radiation is much
greater than the temperature of the CMBR then the spin temperature drops
out of the expression for the emitted energy.  In such a case the energy
$dE$ emitted by a set of hydrogen atoms in an interval $dt_e$ is given by 
\begin{eqnarray}
dE &=& \hbox{Rate of Transition} \times \hbox{Energy carried by a photon}
\nonumber \\
 & & \hbox{ \ } \times \hbox{Number of hydrogen atoms} \times dt_e \nonumber \\
 &=& \frac{3}{4} A_{21} \times h\nu_e \times \frac{M_{H_I}}{m_p} dt_e
\end{eqnarray}
where $m_p$ is the mass of a proton and $M_{H_I}$ is the total mass in
neutral hydrogen.  We will use the mass in neutral hydrogen contributed by
one particle in the N-Body simulation to compute total energy emitted by
each ``N-Body particle''.  The mass $M_{H_I}$ is given by
\begin{eqnarray}
M_{H_I} &=& M_{part} f_b f_n = M_{part} \frac{\Omega_b}{\Omega_0}
f_n \nonumber\\
&=& 8.1 \times 10^9 M_\odot \left(\frac{\Omega_b}{0.06}\right)
\left(\frac{f_n}{0.5}\right)
\end{eqnarray}
here $f_b$ is the fraction of baryons, $\Omega_b$ is the contribution
of baryons to the density parameter and $f_n$ is the neutral
fraction.  We have chosen $\Omega_b=0.06$ as this value compares well
with the observed abundance of light elements and primordial
nucleosynthesis \cite{copnuc}.  Using this, we can estimate the flux
received by an observer from an ``N-Body particle''.  For
$\Omega_{nr}=\Omega_0 =1$ models the flux contributed by one particle
at redshift $z=3.34$ is given by 
\begin{equation}
S_\nu = 1.1 \mu Jy \left(\frac{M_{H_I}}{8.1\times 10^9
M_\odot}\right )\left(\frac{\Delta \nu_0}{175kHz}\right)^{-1}
\end{equation}
The frequency width used here corresponds to a velocity dispersion of
$200$km$/$s. (The corresponding number for the model with
$\Omega_\Lambda = 0.4$ is $0.77\mu Jy$.) 

\begin{figure}
\epsfxsize=3.25 true in \epsfbox[91 84 521 513]{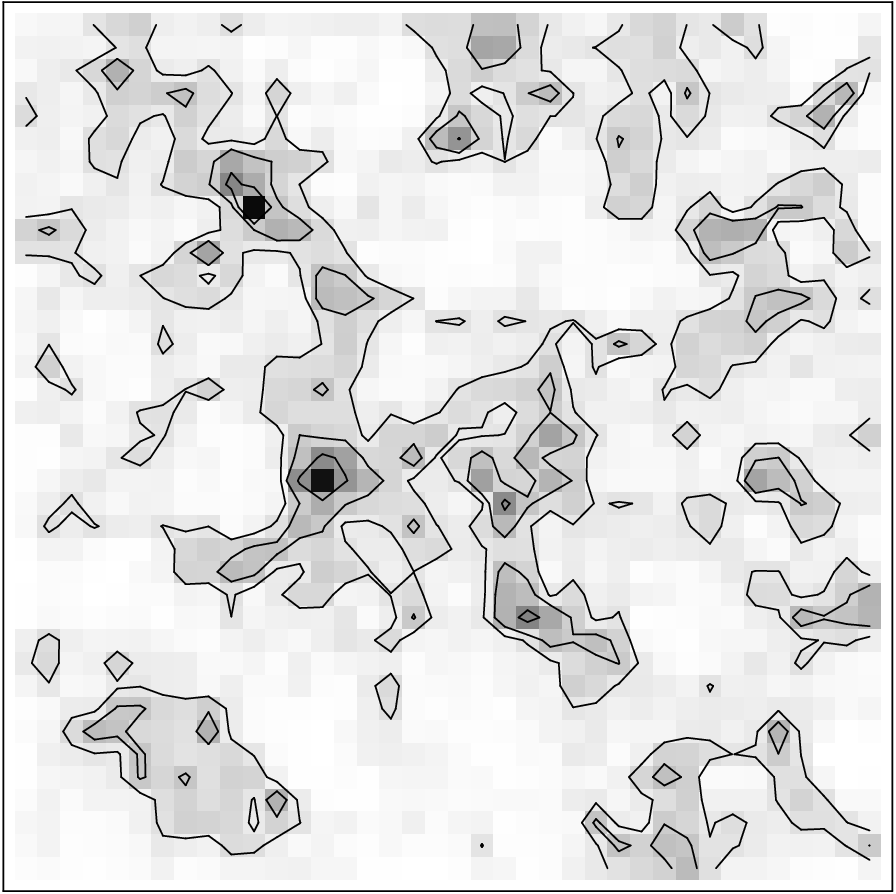}
\caption{This figure shows a sample radio map for model~I (sCDM) at redshift
$z=3.34$.  This map is shown for one frequency channel of width
$125$kHz [corresponding to a velocity width of $115$km$/$s].  Each pixel
corresponds to an angular resolution of $3.2$ 
arc minutes and a physical scale of about $3h^{-1}$Mpc.  The contour
levels correspond to $15$, $30$ and $60\mu$Jy.} 
\end{figure}
\setcounter{figure}{2}

\begin{figure}
\epsfxsize=3.25 true in \epsfbox[91 84 521 513]{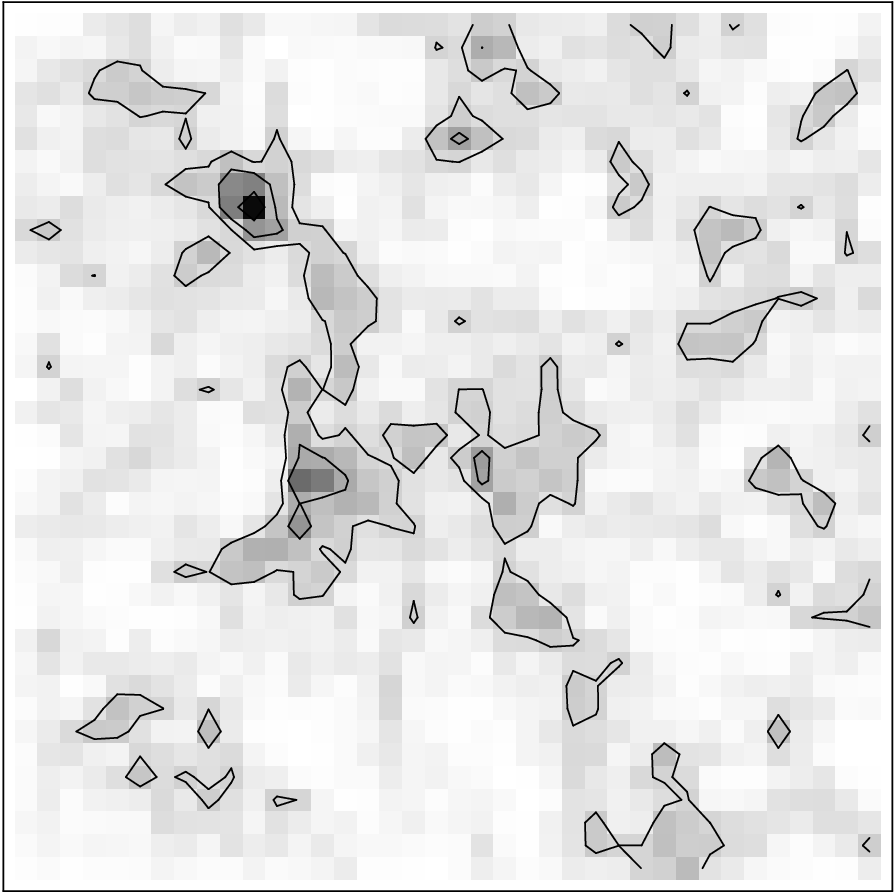}
\caption{Continued. This figure shows a sample map for model II (MDM).}
\end{figure}
\setcounter{figure}{2}

\begin{figure}
\epsfxsize=3.25 true in \epsfbox[91 84 521 513]{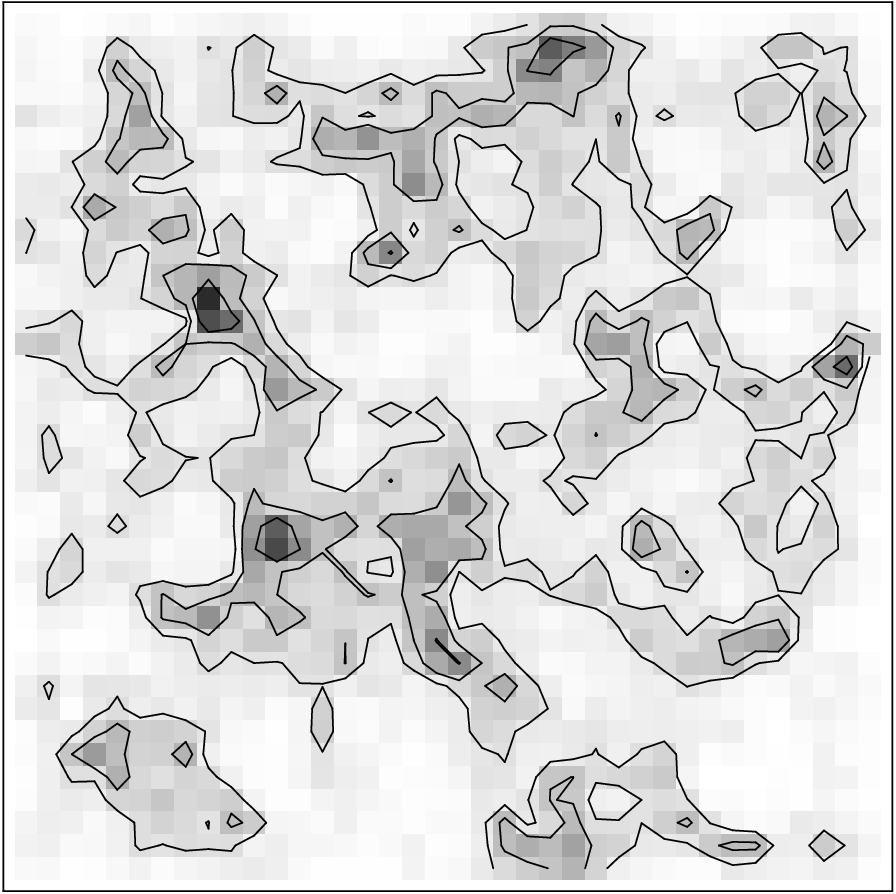}
\caption{Continued. This figure shows a sample map for model III (LCDM).}
\end{figure}

\section{Results}

In this section we shall outline the results of analysis of the radio
maps generated from N-Body simulations.  We begin with a pictorial
preview of the radio maps.  Figure~3 shows a sample radio map for each
of the three models at redshift $z=3.34$.  The panels of this 
figure show one frequency channel (chosen to be $125$kHz; which - in 
velocity units - corresponds to about $115$km$/$s.)  The contours in
these radio maps correspond to $15$, $30$ and $60\mu$Jy.  The pixel
size is $3.2$ arc minutes and it corresponds to a comoving scale of
$2.7h^{-1}$Mpc for models~I and II. ($3.5h^{-1}$Mpc for model~III.)
It is clear from these panels that models~I (sCDM) and III (LCDM) have
comparable signal whereas model~II (MDM), as it has less power at
smaller scales, has somewhat lower signal.  The profile of peaks in
the sCDM model is steeper than that in the LCDM model.

Figure 4 shows similar maps for $z=5.1$ [$\nu_o = 233$MHz].  The width
of one channel in this case corresponds to a velocity width of $161$km$/$s.
Angular size of each pixel is $4.5$ arc minutes and this corresponds to a
comoving scale of $4.7h^{-1}$Mpc for models~I (sCDM) and II(MDM).
($5.4h^{-1}$Mpc for model~III (LCDM).)   These maps clearly show that
there are no small scale 
structures in these, in comparison with the maps at $z=3.34$.  One
reason for the gentler variation in signal from one pixel to another
arises from the fact that, having assumed that the IGM is not ionised,
we have not discarded matter in under dense regions while computing the
signal.  The signal is much higher as the comoving volume enclosed in
each pixel/channel is larger than the corresponding volume at redshift
$3.34$. 

These qualitative features can also be seen in figure~5 that shows a
sample spectrum from the simulations of model~III (LCDM) at these two
redshifts. The spectra are shown as a function of channel number and
the signal is 
shown in $\mu$Jy.  The typical width of high peaks seen in these graphs
are much larger than the velocity dispersion of $200$km$/$s ascribed
to individual particles.

\begin{figure}
\epsfxsize=3.25 true in \epsfbox[91 84 521 513]{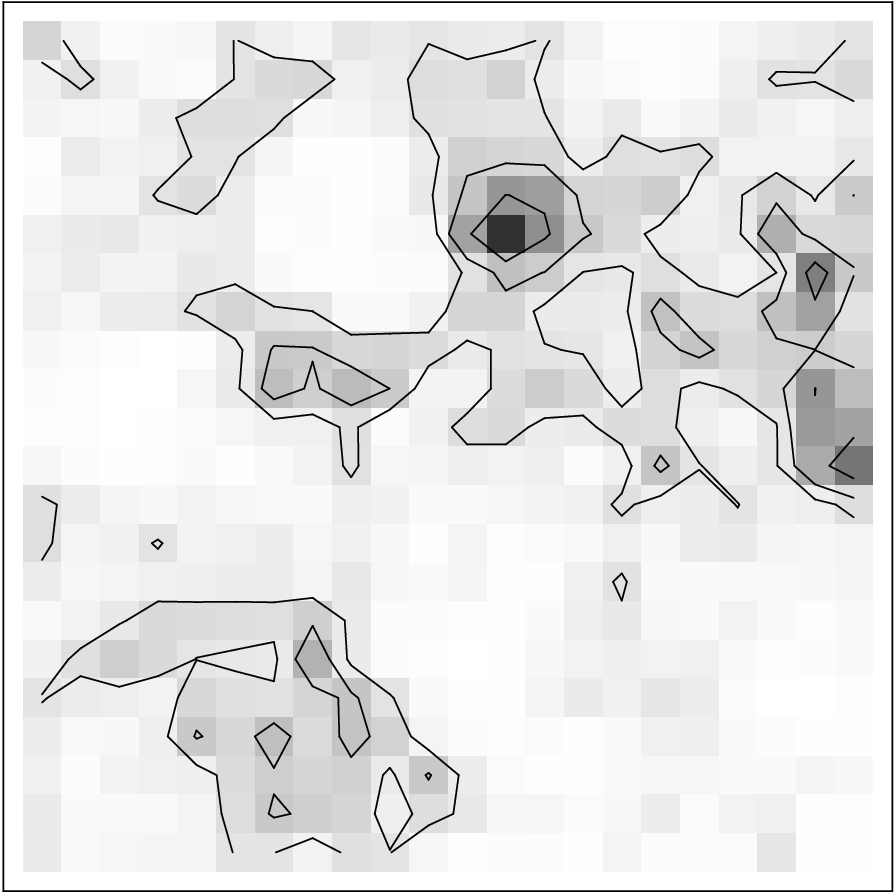}
\caption{This figure shows a sample radio map for model~I (sCDM) at
redshift $z=5.1$.  This map is shown for one frequency channel of width
$125$kHz (corresponding to a velocity width of $160$km$/$s).  Each pixel
corresponds to an angular resolution of $4.5$ 
arc minutes and a physical scale of about $5h^{-1}$Mpc.  The contour
levels correspond to $40$, $80$, $120$ and $200\mu$Jy.} 
\end{figure}
\setcounter{figure}{3}

\begin{figure}
\epsfxsize=3.25 true in \epsfbox[91 84 521 513]{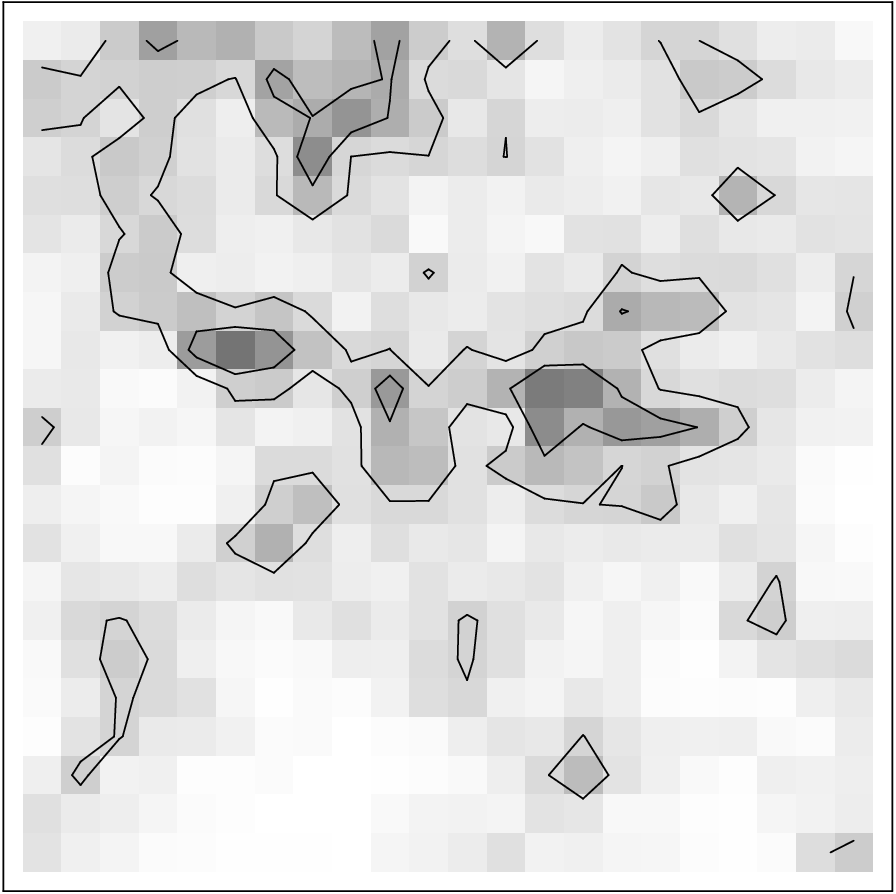}
\caption{Continued. This figure shows a sample map for the MDM model
(model II).} 
\end{figure}
\setcounter{figure}{3}

\begin{figure}
\epsfxsize=3.25 true in \epsfbox[91 84 521 513]{fig4III.ps}
\caption{Continued. This figure shows a sample map for the LCDM model
(model III).} 
\end{figure}

\begin{figure}
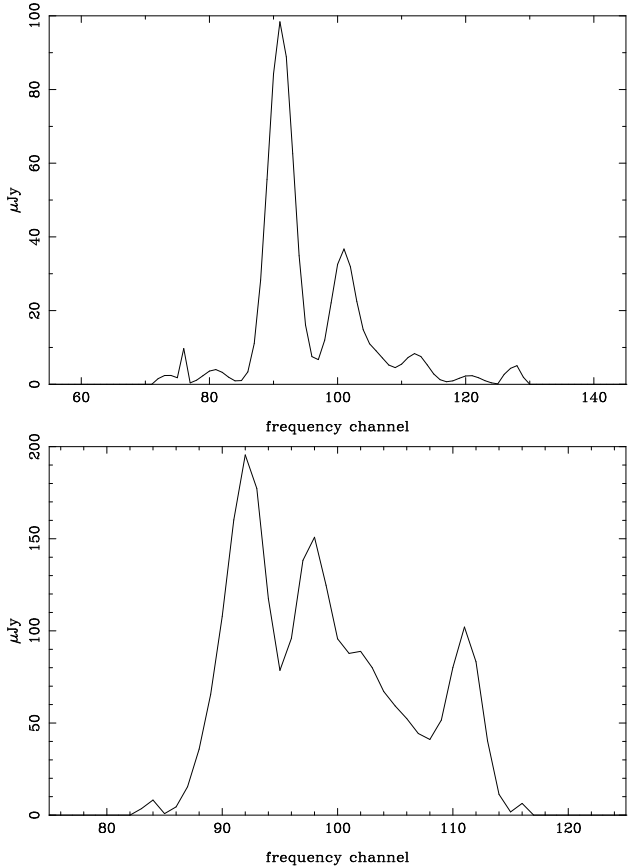

\epsfxsize=3.25true in \epsfbox[38 24 541 371]{fig5.1.ps}
\epsfxsize=3.25true in \epsfbox[38 24 541 371]{fig5.2.ps}
\caption{This figure shows a sample spectrum for each of the redshifts.
The spectra shown here are taken from the simulated radio maps of the
LCDM model (model~III).  The top panel corresponds to $z=3.34$ and the
lower panel to 
$z=5.1$.  The signal in $\mu$Jy is shown as a function of frequency
channel [channel width=$125$KHz].  This figure shows that the signal at
$z=5.1$ has few small scale variations in comparison with the signal at
$z=3.34$.}
\end{figure}

\begin{figure}
\epsfxsize=3.25true in \epsfbox[69 64 529 522]{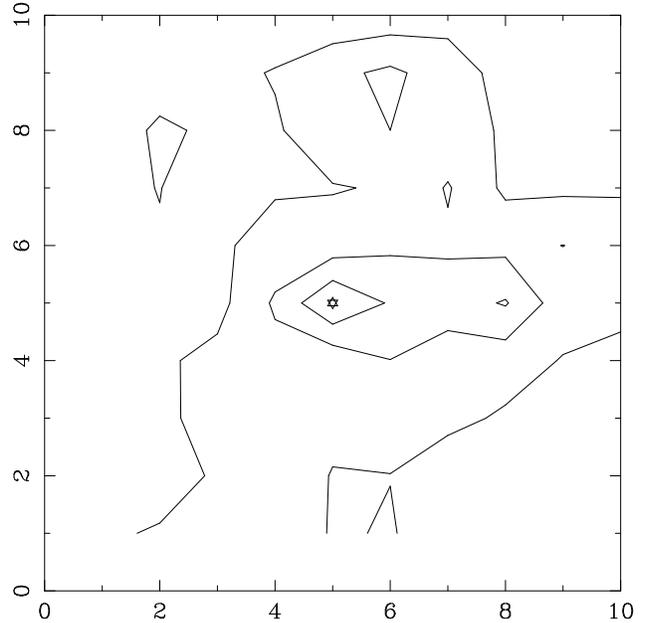}
\caption{This figure shows the contours for $f*S_{max}$ around a peak
for $f=0.75$, $.5$ and $0.25$.
These contours is for one of the highest peaks in simulated maps for
model~III (LCDM).  The peak is located at (5,5) in this picture and is marked
by a star.  It is clear
from this figure that the signal can be averaged over neighbouring pixels
to improve signal to noise ratio.  However the shape of of this contour
suggests that the amount of gain from such an optimisation will be
limited.} 
\end{figure}

In order to make quantitative estimates of the possibility of detecting
neutral hydrogen at high redshifts we computed the amplitude of three
highest peaks in the radio maps for the three models.  Table~1 lists the
amplitudes and the full width at half maximum for these models for the
three highest peaks in each case at redshift $3.34$.  The last column
lists the number of pixels enclosed within the contours of half maximum.
The contours were drawn for signal averaged over frequency channels.  The
number of channels used for averaging was taken to be the FWHM for the
relevant peak.  This number indicates the typical angular size of a flux
peak and the signal can be smoothed at this scale to enhance signal to noise
ratio.  However, the shapes of these peaks are somewhat arbitrary and it
is difficult to suggest a generic smoothing function for improving chances
of detection.  Figure~6 shows contours around one such peak, showing the
level at which the signal drops to $.75$, $.5$ and $.25$ its maximum
value.  This figure demonstrates the point mentioned above regarding
shapes of these contour.

\begin{table}
\begin{tabular}{llll}
Model \ \ & Amplitude ($\mu$Jy) \ \ \ & FWHM (MHz) \ \ &  $n_{HM}$ \\
I & 137.2 & 1.5 & 3 \\
 & 101.5 & 1.0 & 11 \\
 & 99.8 & 0.875 & 7 \\
 & & & \\
II  & 87.1 & 1.0 & 6 \\
 & 97.7 & 0.75 & 8 \\
 & 93.5 & 0.875 & 12 \\
 & & & \\
III & 124.6 & 0.875 & 11 \\
 & 111.3 &  0.875  & 13 \\
 & 103.8 &  1.125  &  8 \\ 
\end{tabular}
\caption{This table lists details of three highest peaks in the simulated
radio maps for the three models of structure formation being considered
here for the redshift window at $z=3.34$ ($327$MHz).  We have listed the
amplitude of peaks in micro Janskys, full width at half maximum in MHz
and the number of pixels enclosed in the contours of half maximum.  It
is clear that models~I (sCDM) and III (LCDM) have a higher signal as
compared to model~II (MDM).}
\end{table}

Even though we cannot suggest the most optimum smoothing function for
improving the signal to noise ratio, we have confirmed that the signal
to noise ratio does improve for
smoothing with a square top hat window of size~$\approx n_{HM}^{1/2}$
where $n_{HM}$ is the number of pixels enclosed within the contour of
$S_{max}/2$.  We assumed that the noise scales as $n^{-1/2}$ for small
$n$.  The signal to noise ratio for models~II (MDM) and III (LCDM) can
be improved by a 
factor two in this manner.  The standard CDM model [model~I] has more
small scale power and hence the peaks are much sharper and therefore the
gain in signal to noise ratio by smoothing is somewhat limited in this
case.

In order to estimate the integration time required for detection and
imaging of these peaks we need to know the root mean square amplitude of
noise expected for the GMRT receivers.  The expected noise for the central
array is
\begin{eqnarray}
\hbox{rms noise} &=& 44 \mu Jy \left(\frac{T_s}{110K}\right) \left(\frac{1
MHz}{\Delta \nu} \right)^{1/2} \left( \frac{100 hrs}{\tau} \right)^{1/2}
\nonumber \\
&=& 100 \mu Jy \left(\frac{T_s}{250K}\right) \left(\frac{1
MHz}{\Delta \nu} \right)^{1/2} \left( \frac{100 hrs}{\tau} \right)^{1/2}
\end{eqnarray}
The system temperature for $327$MHz window is $110$K and the corresponding
number for $233$MHz is $250$K. \cite{swarup84}

It is clear from table~1 that the peak signal expected in models~I
(sCDM) and III (LCDM) is about three times larger than the rms noise
expected at redshift $3.34$.  Further, the width of the peaks in
frequency space is comparable to $1$MHz.  The
signal to noise level can be enhanced further by smoothing over nearby
pixels.  The number of pixels enclosed in the contour of $S_{max}/2$ for
model~III (LCDM) varies between $8$ and $13$ for the three higher
peaks listed in 
table~1.  Therefore an enhancement by a factor two or more can be
obtained by smoothing over nearby pixels.  This implies that a $3\sigma$
detection of such objects should be possible with an integration time of
$50-100$hrs.  The main uncertainty in this result comes from our lack of
knowledge about the fraction of mass in galaxies.  We have, of course,
tried to reduce the uncertainty by using only regions with density
contrast greater than unity for computing the signal.  As mentioned
before, we find that changing the density threshold to $\delta_c=3$
does not change the amplitude of the highest peaks by more than
$15\%$.  However, as the estimated signal and noise have a similar
amplitude, even a small factor may make all the difference.

Detection of neutral hydrogen at high redshift for model~I (sCDM) should also
require similar integration time.  Here the frequency spread of the
highest peak is very large and compensates, to some extent, the
compactness of the peak in angular co-ordinates.  On the other hand
detection of neutral hydrogen in model~II (MDM) is a more difficult proposition
as the peak signal is small compared to that for the other two models.
The integration time required for model~II (MDM) is at least twice as large
as that for models~I (sCDM) and III (LCDM). 

Table 2 lists the amplitude of three highest peaks for the three models
we are using here for the window at redshift $5.1$.  We have also listed
the full width at half maximum for each of these peaks.  These peaks have
a very shallow profile and the number of pixels, enclosed within the region
where the signal is greater than half of the maximum value, can be very
large.  Therefore we can smooth the signal over a few neighbouring
pixels and improve the signal to noise ratio by a significant amount.
Results for different models in this case are:
\begin{itemize}
\item Model~III (LCDM) predicts a signal that is comparable to the rms noise.
Smoothing over nearby pixels and integration for $50-100$hrs should be
sufficient for a $3\sigma$ detection. 

\item  Model~I (sCDM) also predicts signal at the same level as
model~III (LCDM).  

\item  Model II (MDM) has very low signal as compared to models~I
(sCDM) and III (LCDM)
and will require an integration for $100-200$ hours for a $3\sigma$
detection.
\end{itemize}

These numbers, however, must be considered in light of the possibility
of a patchiness in reheating of the IGM.  \cite{tomo}

\begin{table}
\begin{tabular}{llll}
Model \ \ & Amplitude ($\mu$Jy) \ \ \ & FWHM (MHz)\\
I & 199.0 & 0.75 \\
 & 181.4 & 1.0 \\
 & 163.2 & 0.625 \\
 & & & \\
II  & 138.6 & 0.75 \\
 & 115.2 & 0.875 \\
 & 103.8 & 0.75 \\
 & & & \\
III & 221.7 & 0.75 \\
 & 195.6 &  0.875 \\
 & 153.9 &  0.875 \\ 
\end{tabular}
\caption{This table lists details of three highest peaks in the simulated
radio maps for the three models of structure formation being considered
here for the redshift window at $z=5.1$ ($233$MHz).  We have listed the
amplitude of peaks in micro Janskys and full width at half maximum in
MHz.  It is clear that models~I (sCDM) and III (LCDM) have a higher
signal as compared to model~II (MDM).}
\end{table}

\section{Discussion}

Results of numerical simulations suggest that it could be possible to
detect neutral hydrogen at high redshifts, in proto-clusters.
Detection of these objects with the GMRT will require integration over
$50-100$ hours.  The signal to noise ratio for $z=3.34$ is better than
that expected for $z=5.1$ if we choose the factor $f_{gal}$ to be unity.
This is the most uncertain number in our calculation.  If this number
is much smaller than unity then it may be difficult to image
proto-clusters at this redshift.  It may still be possible to ``detect''
neutral hydrogen statistically by doing a $(\Delta T)/T$ type of an
experiment.  In these one looks for excess correlations than
are expected from noise in the instrument.  For example, the emission
from a region of size $10^6(h^{-1}$Mpc$)^3$ - typical size of the
volume probed in one field of view - is around $370mJy$.  [This number
is obtained by using the average flux in the simulated radio maps.]  
This will appear as excess noise if the individual structures can not
be imaged.  Integration over more than one field of view can
also be used to enhance the chances of detection as one may be able to
pick up a rare density peak \cite{nhI}.

As mentioned above, the biggest uncertainty in our results for
$z=3.34$ is introduced by our lack of knowledge about $f_{gal}$.  This
uncertainty can only be removed with large simulations that can
resolve masses less than $10^{10}M_\odot$ and still cover a volume that
is comparable with that covered by one field of view for the GMRT.
The other option is to carry out a series of N-Body experiments at
different scales for each of the models and deduce $f_{gal}$ for each
model from the ensemble of simulations.

A comparison with earlier estimates for the expected signal
\cite{nhline} shows that the highest signal obtained in simulations is
smaller than that expected from a proto-cluster of mass
$10^{15}M_\odot$.  This could result due to three reasons
\begin{itemize} 
\item We do not have a sufficiently high peak in our realisation of 
the density field.

\item The angular extent of the proto-clusters is considerably larger than 
one pixel.  

\item The velocity dispersion of these clusters is larger than expected.
\end{itemize}
We feel that all of these factors have contributed to make our
estimates smaller than the earlier estimates by a factor $2-5$.

The possibility of detection of neutral hydrogen at redshift $5.1$ may be
easier if the universe has been reheated but not fully reionised.  Another
good feature of $21$cm emission from this epoch is that it only has smooth
large scale variations in signal and hence it should be possible to
enhance the signal to noise ratio by at least a factor $2-3$ by smoothing
over nearby pixels.  However, the results for this epoch are somewhat
uncertain due to our poor knowledge of the physical conditions at
$z>5$.

\section*{ACKNOWLEDGEMENT}

Authors thank K.Subramanian and J.Chenglur for useful discussions.  
JSB thanks CSIR India for financial support.

\label{lastpage}


\begin{thebibliography}{}

\bibitem[\protect\citename{Bagla, Padmanabhan and Narlikar} 1996]{comments} 
Bagla J.S., Padmanabhan T. and Narlikar J.V., 1996, Comments on Astrophysics, 
18, 275 

\bibitem[\protect\citename{Copi, Schramm and Turner} 1995]{copnuc} Copi
C.J., Schramm D.N. and Turner M.S., 1995, Science, 267, 192

\bibitem[\protect\citename{Fall and Rees} 1985]{falrees} Fall S.M. and
Rees M.J., 1985, ApJ, 298, 18

\bibitem[\protect\citename{Giallongo E. et al.} 1994]{gptest}
Giallongo E. et al., 1994, ApJ, 425, L1 

\bibitem[\protect\citename{Kauffman, White and Guideroni} 1993]{kwg}
Kauffman G., White S.D.M. and Guideroni B., 1993, MNRAS, 264, 201

\bibitem[\protect\citename{Kumar, Padmanabhan and Subramanian} 
1995]{nhline} Kumar A., Padmanabhan T. and Subramanian K., 1995,
MNRAS, 272, 544

\bibitem[\protect\citename{Madau, Meiksin and Rees} 1997]{tomo} Madau
P., Meiksin A. and Rees M.J., 1997, ApJ, 475, 429

\bibitem[\protect\citename{Press and Schechter} 1975]{ps75} Press
W.H. \& Schechter P., 1975, ApJ, 187, 452

\bibitem[\protect\citename{Scott and Rees} 1990]{screes} Scott D. and
Rees M.J., 1990, MNRAS, 247, 510

\bibitem[\protect\citename{Subramanian and Padmanabhan} 1993]{nhI}
Subramanian K. and Padmanabhan T., 1993, MNRAS, 265, 101

\bibitem[\protect\citename{Sunyaev and Zeldovich} 1972]{sznh1} Sunyaev
R.A. and Zeldovich Ya.B., 1972, A{\&}A, 20, 189

\bibitem[\protect\citename{Sunyaev and Zeldovich} 1975]{sznh2} Sunyaev
R.A. and Zeldovich Ya.B., 1975, MNRAS, 171, 375

\bibitem[\protect\citename{Swarup } 1984]{swarup84} Swarup G., 1984,
Giant Meter-Wavelength Radio Telescope - Proposal, Radio Astronomy
Centre, TIFR, India.

\bibitem[\protect\citename{van de Hulst} 1945]{cm21} van de Hulst,
H.C., 1945, Nederlandsch Tid voor Naturkunde, 11, 201

\bibitem[\protect\citename{Weinberg et al.} 1996]{davwein} Weinberg D.H.,
Hernquist L., Katz N.S. and Miralda-Escud\'e J., 1996, {\it Cold Gas at
High Redshift} eds. M.~Bremer, H.~Rottgering, C.~Carilli and P.~van de Werf, 
Kluwer, Dordrecht.

\bibitem[\protect\citename{White, Efstathiou and Frenk} 1992]{sigma8}
White S.D.M., Efstathiou G. and Frenk C.S., 1992, MNRAS, 262, 1023 

\bibitem[\protect\citename{White and Frenk} 1991]{wf91} White S.D.M. and
Frenk C.S., 1991, ApJ, 379, 52

\bibitem[\protect\citename{Wieringa, de Bruyn and
Katgert} 1992]{nh1obsv} Wieringa M.H., de Bruyn A.G. and Katgert
P., 1992, A\&{A}, 256, 331 

\bibitem[\protect\citename{Wolfe et al.} 1995]{dlas_omnh1} Wolfe A.M.,
Lanzetta K.M., Foltz C.B. and Chaffee F.H., 1995, ApJ, 454, 698

\bibitem[\protect\citename{Wright et al.} 1996]{cobe4} Wright E.L., 
Bennet C.L., Gorski K., Hinshaw G. and Smoot G.F., 1996, ApJ, 464, L21

\end{thebibliography}
\end{document}